\documentclass[aps,prx,reprint,superscriptaddress]{revtex4-2}

\usepackage[dvipsnames]{xcolor}
\usepackage{soul}
\usepackage{ulem}
\usepackage{graphicx}
\usepackage{amsmath}

\makeatletter
\renewcommand\frontmatter@abstractwidth{\dimexpr\textwidth-0in\relax}
\makeatother

\begin{document}

\title{The measurement of Navier slip on individual nanoparticles in liquid}

\author{Jesse F. Collis}
\thanks{These authors contributed equally to the work.}
\affiliation{ARC Centre of Excellence in Exciton Science, School of Mathematics and Statistics, The University of Melbourne, Victoria 3010, Australia}
\author{Selim Olcum}
\thanks{These authors contributed equally to the work.}
\affiliation{Koch Institute for Integrative Cancer Research, Massachusetts Institute of Technology, Cambridge, MA 02139, USA}
\author{Debadi Chakraborty}
\affiliation{ARC Centre of Excellence in Exciton Science, School of Mathematics and Statistics, The University of Melbourne, Victoria 3010, Australia}
\author{Scott R. Manalis}
\affiliation{Koch Institute for Integrative Cancer Research, Massachusetts Institute of Technology, Cambridge, MA 02139, USA}
\author{John E. Sader}
\email[Corresponding author. Email: ]{jsader@unimelb.edu.au}
\affiliation{ARC Centre of Excellence in Exciton Science, School of Mathematics and Statistics, The University of Melbourne, Victoria 3010, Australia}

\date{\today}

\begin{abstract}
\noindent 
The Navier slip condition describes the motion of a liquid, relative to a neighboring solid surface, with its characteristic Navier slip length being a constitutive property of the solid-liquid interface. Measurement of this slip length is complicated by its small magnitude, expected in the nanometer range based on molecular simulations. Here, we report an experimental technique that interrogates the Navier slip length on individual nanoparticles immersed in liquid, with sub-nanometer precision. Proof-of-principle experiments on individual, citrate-stabilized, gold nanoparticles in water give a constant slip length of 2.7$\boldsymbol{\pm}$0.6 nm (95\%~C.I.)---independent of particle size. Achieving this feature of size independence is central to any measurement of this constitutive property, which is facilitated through the use of individual particles of varying radii. This demonstration motivates studies that can now validate the wealth of existing molecular simulation data on slip.
\end{abstract}

\maketitle

\newpage

\section{Introduction}
The interaction of a liquid flowing past a solid surface has been studied extensively for the last century. It is now well accepted that the traditional no-slip boundary condition normally holds at large, macroscopic scales. Yet, use of the no-slip condition  is being restricted with the increasing miniaturization of technology. Tremendous progress has been made in the last few decades towards understanding this situation using both molecular dynamics simulations \cite{thompson1997,barrat1999,huang2008,voronov2008} and high resolution experimental techniques~\cite{neto2005,whitby2007,lauga2007,bocquet2010,shu2017,bocquet2020}.

\begin{figure*}
\includegraphics[width=0.85\textwidth]{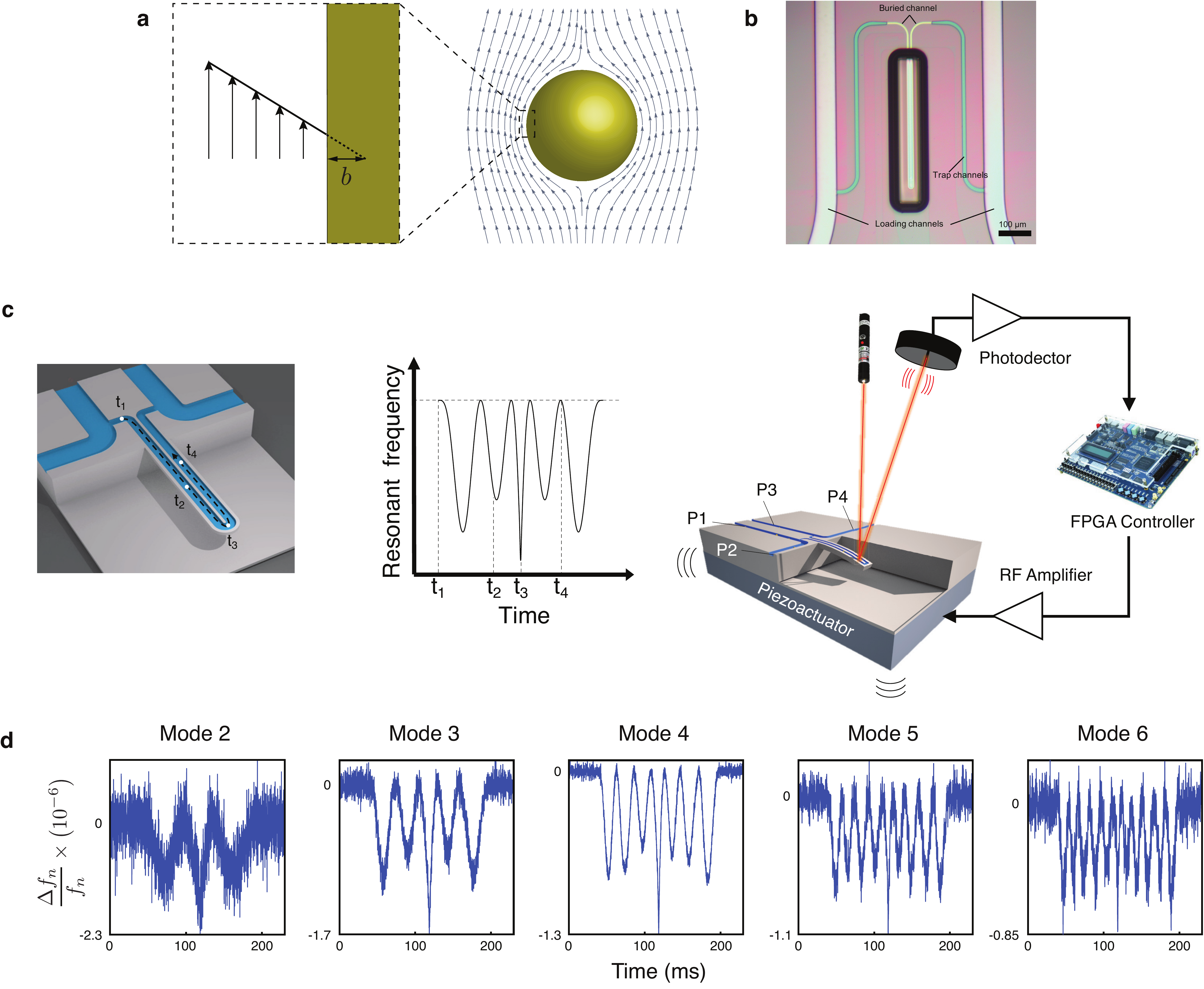}
  \caption{ \rm Schematic of SMR measurement protocol.  \textbf{a}, Streamlines of a liquid flowing past a sphere, which is moving with steady velocity at zero Reynolds number (inertialess flow). Geometrically, the `slip length' is the extrapolated distance into the solid boundary where the no-slip boundary condition would apply (this interpretation holds exactly for a flat surface); see equation~(\ref{eq:navier}). \textbf{b}, Optical micrograph of the SMR showing the cantilever enclosed in a vacuum chamber with an embedded microfluidic channel that is connected via loading, trap and buried channels; nanoparticles are passed through these channels for measurement, one particle at a time. \textbf{c}, [left] Cross section of the SMR showing a U-shaped fluid channel embedded to the interior of the resonating cantilever. The dotted line represents the transit of a nanoparticle flowing through the SMR. The illustration shows time lapse snapshots of a single particle as it passes through the channel at times, $t_1 < t_2 < t_3 < t_4$. [middle]  Each negatively buoyant nanoparticle increases the inertial mass of the SMR, reducing the natural resonant frequencies of the SMR's flexural modes and therefore traces out the mode shapes (squared) of the SMR. A theoretical calculation of the resonant frequency of mode 3 versus time is shown here; the times, $t_1$ to $t_4$, correspond to particle positions labelled in the left subfigure. The magnitude of resonant frequency change depends both on the buoyant mass of the nanoparticle and the boundary condition at the nanoparticle surface; see equations~(\ref{eq:freqshiftformula})--(\ref{eq:alphaFull}). [right] An optical lever setup is used to convert the SMR tip motion to an electrical signal.  An FPGA uses this signal to determine the drive signals for each of the modes using an array of digital phase-locked loop controllers. These drive signals are superposed and used to excite a piezo-ceramic actuator attached to the SMR chip. \textbf{d}, Real measured (sample) data are shown for the resonant frequencies of each vibrational mode as a 126 nm radius gold nanoparticle passes through the SMR. A sample of data versus nanoparticle size is given in Supplementary Fig.~4.
}
\label{figschematic}
\end{figure*}

Deviation from the no-slip boundary condition is described by the Navier slip condition~\cite{maxwell1879,vincenti1967,cercignani1969},
\begin{equation}
\left[ \left(\mathbf{u} - \mathbf{u}_s - b \, \mathbf{n} \cdot \mathbf{S} \right)\cdot\left(\mathbf{I} - \mathbf{nn}\right) \right]_\mathrm{surface} =  \mathbf{0},
\label{eq:navier}
\end{equation}
where $\mathbf{n}$ is the unit normal to the surface (into the liquid), $\mathbf{I}$ is the identity tensor, $\mathbf{u}$ and $\mathbf{u}_s$ are the liquid and solid velocity vectors, respectively, $\mathbf{S}\equiv 2\mathbf{e} = \nabla\mathbf{u}+\left(\nabla\mathbf{u}\right)^\mathrm{T}$ where $\mathbf{e}$ is the rate-of-strain tensor in the liquid and $b$ is a proportionality constant: the `slip length'. For a flat surface, $b$ quantifies the depth of a hypothetical surface beneath the physical one where the no-slip condition would hold; see Fig.~\ref{figschematic}a. Importantly, the Navier slip length is a constitutive property of the liquid-solid interface, i.e., it is independent of the flow geometry, and its size.

Theoretical studies of slip in liquid are largely based on molecular dynamics simulations~\cite{thompson1997,barrat1999,huang2008,voronov2008} and arguments from linear response theory~\cite{bocquet2013,huang2014}. There have also been suggestions that slip at the liquid-solid interface can be modeled as a kinetic rate process~\cite{lichter2007,wang2019}. The key insight to come from molecular simulations is that, at low shear rates, the Navier slip condition emerges which depends sensitively on the properties of the interface~\cite{bocquet2010}. Importantly, the wettability of the interface is observed in both molecular dynamics and linear response theory studies to affect the slip length. Interfaces with high wettability usually produce slip lengths of a few nanometers or less; conversely, interfaces with low wettability can have slip lengths in the tens of nanometers~\cite{huang2008}. Given the slip length is a constitutive property of the liquid-solid interface (discussed above), these small slip lengths establish that slip is normally unimportant for flows at normal macroscopic length scales. That is, the no-slip condition can be used with confidence, apart from some important exceptions such as the moving three-phase contact line~\cite{snoeijer2013}.

On the experimental front, the degree of slip at the liquid-solid interface has been interrogated using a range of approaches including monitoring thermal fluctuations of free surfaces~\cite{pottier2015}, thermal motion of particle suspensions~\cite{joly2006}, an array of particle velocimetry techniques~\cite{bouzigues2008,lasne2008,li2010,li2015}, atomic force microscope methodologies~\cite{bonaccurso2003,honig2007,maali2008,guriyanova2009,henry2009} through to use of the surface forces apparatus \cite{cottin2008}. Large slip lengths ($>$$1\, \mu$m) have mostly been attributed to dissolved gases on the solid surface \cite{de2002,lauga2007,shu2017}.  For a review of nanofluidics, experiments used to measure slip at the liquid-solid interface and their findings, see Refs.~\cite{neto2005,whitby2007,zhu2012,lauga2007,bocquet2010,shu2017,bocquet2020}.

Interestingly, confinement has also been observed to modify slip, with large slip lengths measured for liquid flows  confined within carbon nanotubes~\cite{majumder2005,whitby2007,radha2016,secchi2016}. Secchi~\textit{et al.}~\cite{secchi2016} reported a strong and systematic dependence of the measured slip length on tube diameter: $b=300$ nm for a 15 nm radius tube whereas $b=17$ nm for a (larger) 50 nm tube---in qualitative agreement with previous molecular simulations~\cite{thomas2009,kannam2013}. This shows that confinement can modify the above-mentioned constitutive nature of the slip length. The reported dependence is also affected by electrochemical properties of the liquid-solid interface, with no-slip being observed for boron nanotubes at all tube radii~\cite{secchi2016}. 

Here, we develop an experimental technique to measure the Navier slip length on individual nanoparticles in liquid, with sub-nanometer precision, that does not involve confinement. This is achieved using a new modality of suspended microchannel resonators (SMRs)~\cite{burg2007,dohn2007,lee2010,olcum2014,olcum2015,yan2017} that interrogates the hydrodynamic flow generated by a single nanoparticle in an unconfined liquid. The use of nanoparticles of different size enables the constitutive nature of the slip length to be examined and confirmed. This critical step ensures robust determination of the slip length, that can be compared to the vast library of molecular simulations of slip~\cite{thompson1997,barrat1999,huang2008,voronov2008}. The developed technique is demonstrated in a proof-of-principle measurement using citrate-stabilized gold nanoparticles in water. 
This measurement is compared to available molecular simulations.

\section{Experimental approach}

\subsection{Measurement protocol}\label{secMeas}

SMRs are cantilevered inertial mass sensors that allow ultra-sensitive measurements---at the attogram level~\cite{olcum2014}---in liquid environments. This feature is central to the measurement of slip. An optical micrograph of the SMR used in this work is provided in Fig.~\ref{figschematic}b, showing its (sensing) embedded microfluidic channel, which is fed via loading, trap and buried channels; see Appendices~\ref{Measurement setup} and \ref{Design of the suspended microchannnel resonator} for further details. The embedded, `U-shaped' channel is completely enclosed by the cantilever, which allows  particles to flow through the device; as illustrated in Fig.~\ref{figschematic}c. As a negatively buoyant, individual nanoparticle flows through the embedded channel of the SMR, it increases the inertial mass of the sensor. This reduces the natural resonant frequencies of the SMR's vibrational modes; Figs.~\ref{figschematic}c and \ref{figschematic}d show a theoretical calculation for mode 3 and actual measurements for modes 2 to 6, respectively. The nanoparticles remain suspended and do not sediment, due to the action of Brownian forces.

\begin{figure}[!ht]
\centering
\includegraphics[width=\columnwidth]{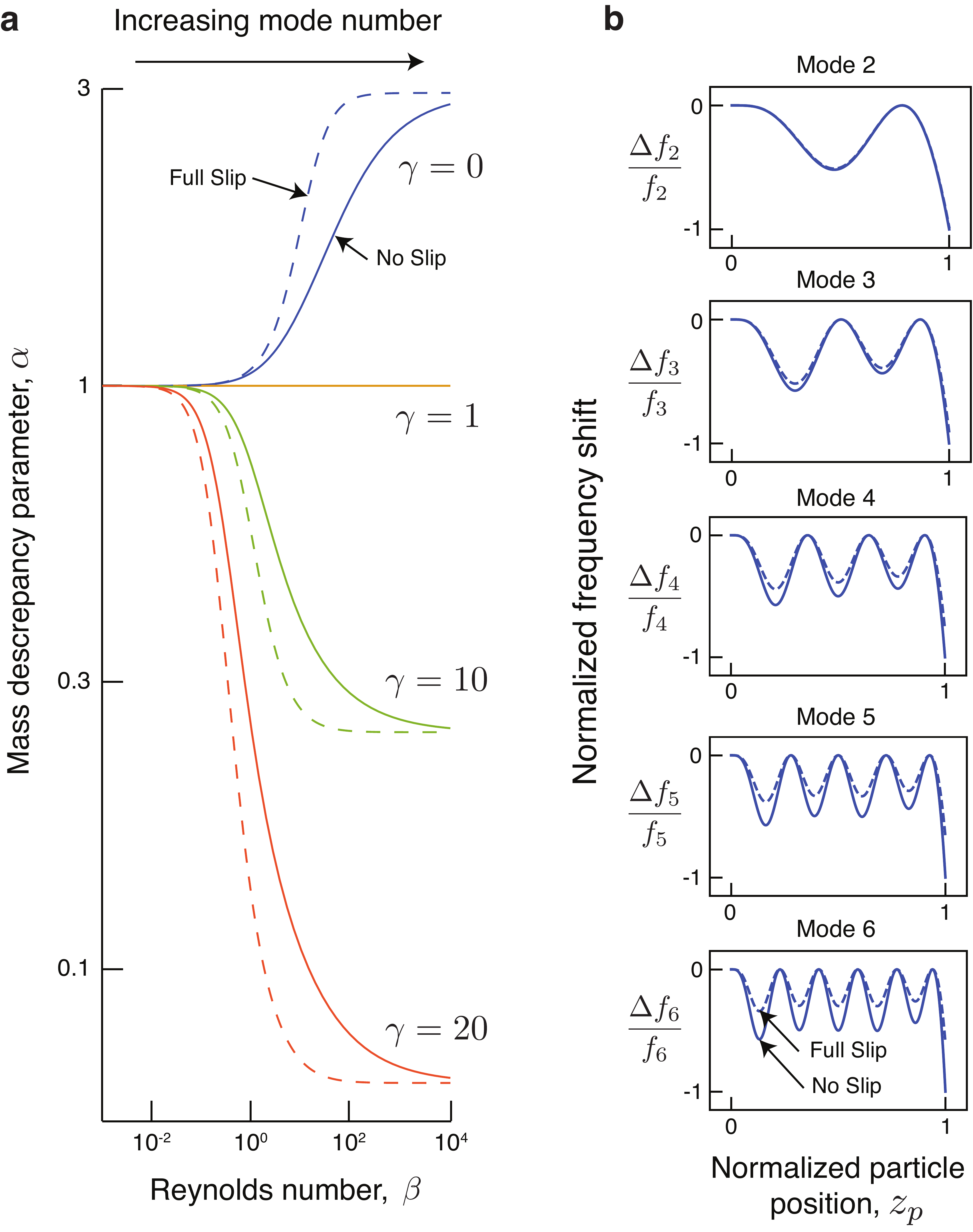}
  \caption{ \textbf{\rm Effect of buoyant density ratio and slip length on the SMR frequency.  \textbf{a}, Mass discrepancy parameter, $\alpha$, defined in equation~(\ref{eq:alphaformula}), for a nanoparticle whose radius is far smaller than the SMR channel dimensions, as a function of the density ratio, $\gamma$, and oscillatory Reynolds number, $\beta$. No-slip at the particle's surface coincides with $\lambda \equiv b/R =0$, (solid), whereas full slip, i.e., zero shear stress, at the particle's surface is given by $\lambda\rightarrow\infty$, (dashed). Equation (\ref{eq:alphaFull}) gives the analytical formula for $\alpha$. In general, increasing slip reduces the effective size of the particle (by reducing drag), thereby reducing the observed buoyant mass. This explains the variation between the `No slip' and the `Full Slip' curves. \textbf{b}, Theoretical frequency shifts for a particle with a density ratio of $\gamma = 20$ and an oscillatory Reynolds number for mode 2 of $\beta = 0.03$ (this corresponds to an $\approx$150 nm radius gold nanoparticle in the present experiments). Frequency shifts are normalized by the maximum absolute value of the no-slip curve. Motion relative to the SMR is enhanced with increasing mode number (and hence frequency), causing a decrease in the observed buoyant mass of the particle (for particles more dense than the liquid). Slip increases this relative motion, causing a further decrease in the observed buoyant mass.}}
\label{figslipDeviation}
\end{figure}

Previous SMR measurements assume the particles move in concert with the surrounding liquid, enabling the true buoyant mass of the particle to be extracted. However, when buoyancy forces acting on the particle become large (by, for example, a large density imbalance between the particle and the liquid), the particle can move relative to the surrounding liquid, and the observed buoyant mass of the particle is no longer equal to its true buoyant mass~\cite{yan2017}. In the present experiments, this motion is induced by using gold nanoparticles in water and by exciting the SMR's higher-order flexural modes, whose resonant frequencies increase with mode order. Because this relative motion must drive a flow in the liquid, it is intimately connected to the hydrodynamic boundary condition at the particle's surface. This connection enables measurement of the Navier slip length through mass measurements on nanoparticles of different size.

Each individual nanoparticle is suspended in water and flows through the microfluidic channel of the SMR, which is simultaneously driven to oscillate at multiple vibrational modes spanning more than three octaves in frequency. A commensurate theoretical and statistical framework is developed enabling the interpretation of these measurements on individual nanoparticles. Hundreds of measurements on each nanoparticle are collected---by trapping the particle and flowing it back and forth through the SMR---to mitigate the inherent frequency noise in the experiments, and thus accurately and precisely determine the slip length of each nanoparticle. Thousands of measurements are performed in total (Appendices~\ref{Observed buoyant particle mass using each SMR mode} and \ref{Selecting appropriate gold nanoparticles}).

\subsection{Theoretical framework}\label{secTheory}

Mass measurements using inertial sensors relate the measured frequency shift of each vibration mode to an observed buoyant particle mass,
\begin{equation}
\frac{\Delta f_n}{f_n} = -\frac{1}{2} \frac{M^\mathrm{obs}_\mathrm{p}}{M_\mathrm{SMR}} U_n^2 (z_\mathrm{p}),
\label{eq:freqshiftformula}
\end{equation}
where $\omega_n = 2 \pi f_n$ is the angular resonant frequency of the $n^\mathrm{th}$ mode of the SMR without the particle, $\Delta f_n$ is the resultant frequency shift in the presence of the particle, $M^\mathrm{obs}_\mathrm{p}$ and $M_\mathrm{SMR}$ are the observed buoyant  mass of the particle and the mass of the SMR, respectively, $U_n (z_\mathrm{p})$ is the scaled displacement mode shape of the SMR at the particle position, $z=z_\mathrm{p}$, under the normalization $\int_0^L U^2_n\left(z\right)dz = L$, where $L$ is the SMR's length; equation~(\ref{eq:freqshiftformula}) holds provided $M^\mathrm{obs}_\mathrm{p} \ll M_\mathrm{SMR}$. Particle motion relative to the SMR modifies the inertial mass of the system so that the observed buoyant particle mass is not the true buoyant mass of the particle. Use of equation~(\ref{eq:freqshiftformula}) in such measurements then leads to a discrepancy relative to the true buoyant particle mass, which is quantified by the parameter,
\begin{equation}
\alpha \equiv \frac{M_\mathrm{p}^\mathrm{obs}}{M^\mathrm{true}_\mathrm{p}},
\label{eq:alphaformula}
\end{equation}
where `obs' and `true' refer to the observed buoyant mass of the particle, calculated by application of equation~(\ref{eq:freqshiftformula}), and the true buoyant mass of the particle, respectively. If there is no motion of the particle relative to the SMR, then $\alpha = 1$ and the true buoyant mass of the particle is recovered from equation (\ref{eq:freqshiftformula}).

An analytical formula is derived for $\alpha$ under the following assumptions: (i) the particle is a solid sphere, (ii) it is located far from any internal walls, (iii) its radius, $R$, is much smaller than the SMR length, $L$, and (iv) the particle oscillation amplitude is small relative to $R$. All assumptions are satisfied in the measurement, which is insensitive to particle non-sphericity (Appendix~\ref{Effect of surface roughness on slip length measurements}); analysis of the error induced by wall bounded flows is in Appendix~\ref{Effect of the SMR walls on slip length measurements}. Full details of the calculation are provided in Supplementary section 1, and the key result is
\begin{equation}
\label{eq:alphaFull}
\alpha= \frac{F(2+\gamma)+3 \left(1+2\gamma\right) G}{F(1+2\gamma)+\left(1+2\gamma\right)^2 G},
\end{equation}
where
\begin{subequations}
\begin{align}
F(x) &= 81 \left(1+2\lambda\right)^2 \!\left(1+\sqrt{2\beta}+\beta+\frac{\sqrt{2} \beta^{3/2} x}{9}\right), \label{eq:massFactor_coeff1} \\
G &= \left(1+3\lambda\right)^2\beta^2+\sqrt{2}\lambda\left(1+3\lambda\right)\beta^{5/2}+\lambda^2\beta^3. \label{eq:massFactor_coeff2}
\end{align}
\end{subequations}
Here, $\beta \equiv \omega_n R^2 \rho_\mathrm{f}/\mu$ is the oscillatory Reynolds number and $\gamma \equiv \rho_\mathrm{p}/\rho_\mathrm{f}$ is the particle-to-liquid density ratio, where $R$ is the particle radius, $\rho_\mathrm{p}$ is the particle density, $\rho_\mathrm{f}$ is the liquid density and $\mu$ is the liquid's shear viscosity. The non-dimensional slip length, $\lambda \equiv b/R$, is scaled by the particle radius. The combination of equations~(\ref{eq:freqshiftformula})--(\ref{eq:alphaFull}) fully connect the measured frequency shift to (i) the particle radius, and (ii) the slip length---the primary variables to be measured.

Figure~\ref{figslipDeviation}a provides theoretical results for the mass discrepancy parameter, $\alpha$, as a function of the oscillatory Reynolds number, $\beta$,  for a range of density ratios, $\gamma$. Bounding values for the non-dimensional slip length, $\lambda \rightarrow 0$ and $\infty$, are also given, corresponding to no-slip and zero shear stress (full slip) at the solid surface, respectively; see equation (\ref{eq:navier}). In the limit where the particle density approaches that of the liquid, i.e., $\gamma \rightarrow 1$, the true buoyant particle mass is recovered with $\alpha = 1$. Figure~\ref{figslipDeviation}a also reveals that slip can only affect the measurement for intermediate values of $\beta$, a key requirement in designing the present measurement of slip. 

Figure~\ref{figslipDeviation}b shows theoretical calculations of the frequency shift for each vibration mode of the SMR used in this study. This highlights the effects of slip for a 150 nm gold nanoparticle passing through the SMR. By fitting such theoretical results to the corresponding measured frequency shift curves (as in Fig.~\ref{figschematic}d), the radius and slip length of each nanoparticle can be determined; the required procedure is detailed in \S \ref{Measurements on single nanoparticles}.

\section{Results and Discussion}

We now report the results of  proof-of-principle measurements  demonstrating the proposed method to measure the Navier slip length. This includes a detailed analysis of the measurements and the required data processing to extract the Navier slip length of individual nanoparticles.

\subsection{Measurements on single nanoparticles}\label{Measurements on single nanoparticles}

Gold nanoparticles with identical surface treatment, but different size are used (Appendix~\ref{Selecting appropriate gold nanoparticles}). All nanoparticles are chosen to span the above-mentioned intermediate $\beta$-range, using SMR vibrational modes 2 to 6.  On each pass through the SMR, a frequency shift versus particle position curve is measured for each of the 5 SMR vibrational modes. Fitting each of these curves to equation~\eqref{eq:freqshiftformula} gives an observed buoyant mass, one for each of the excited vibrational SMR modes (Appendix~\ref{Observed buoyant particle mass using each SMR mode}). 

\begin{figure*}
\includegraphics[width=\textwidth]{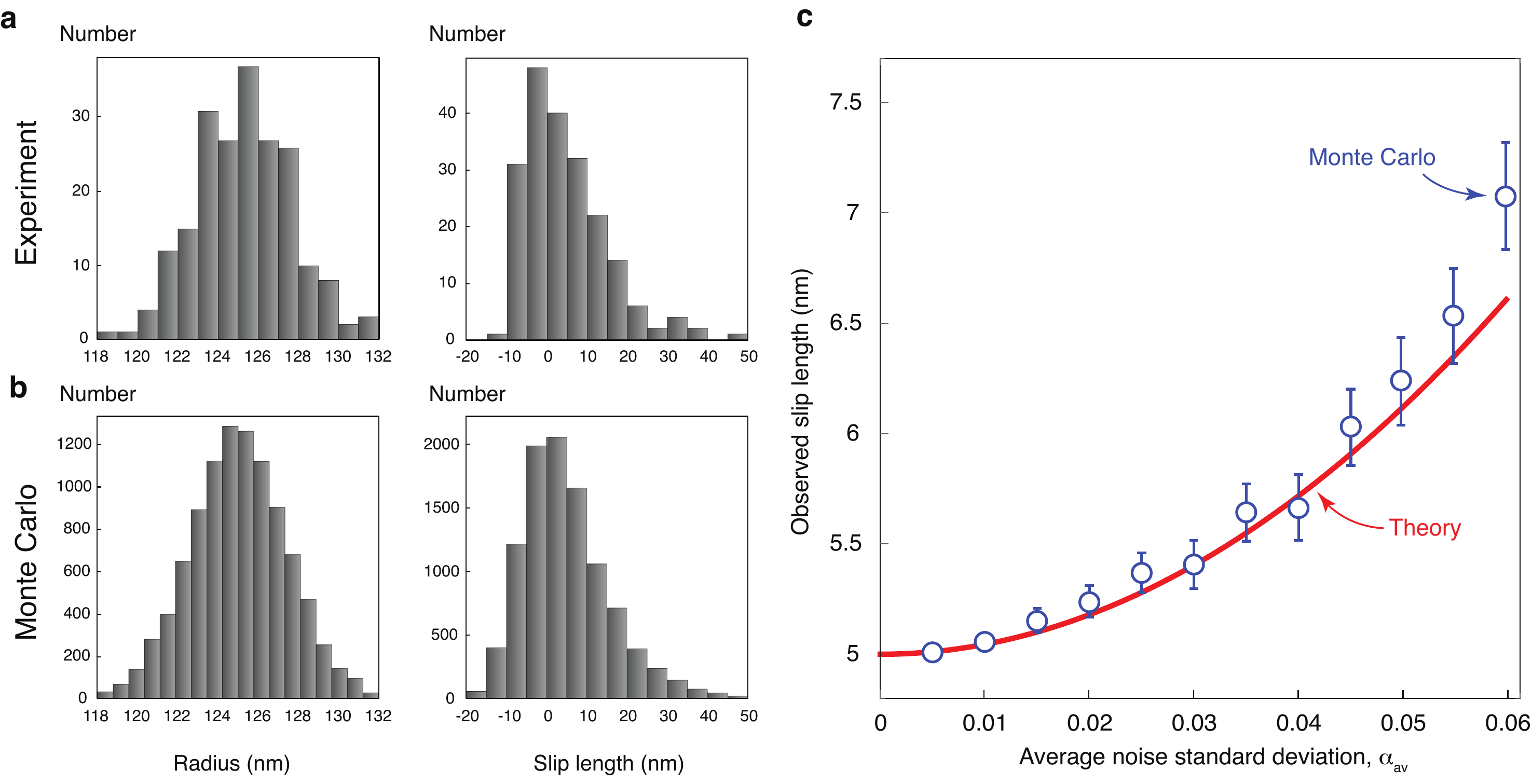}
  \caption{ \rm Histograms of particle radius and slip length, and effect of frequency noise.  \textbf{a}, \textit{Measured} radii and slip lengths obtained from experiments using an SMR with a single gold nanoparticle (204 individual measurements). \textbf{b}, \textit{Simulated} radii and slip lengths for a 125 nm radius gold nanoparticle with a nominal slip length of 5 nm. This differs slightly from Fig.~\ref{figexperiment}, however varying this value has no qualitative effect on the simulated histograms (Appendix~\ref{Monte Carlo simulations}). Comparison to the measured histograms in (a) demonstrates that skewness in the fitted slip length distributions is a result of frequency noise; no frequency noise produces delta functions. \textbf{c}, \textit{Simulated} bias in the mean slip length as the  frequency noise is varied in the system described in (b) (Appendix~\ref{Monte Carlo simulations}). The horizontal axis specifies the average of the frequency noise over all modes (standard deviation). This coincides with the standard deviation of the mass discrepancy parameter, $\alpha_\mathrm{av}$, because $\Delta f_n/f_n$ and $\alpha$ are related linearly; see equations~(\ref{eq:freqshiftformula}) and (\ref{eq:alphaformula}). Measurements in (a) yield $\alpha_\mathrm{av} = 0.0401$. Solid (red) curve gives the theoretical mean of the slip length histogram as a function of frequency noise, using the derived asymptotic theory; see Supplementary section 2. Dots (blue) are Monte Carlo simulations described in (b), with error bars specifying a 95\% C.I.
}
\label{fighistograms}
\end{figure*}

These 5 measurements of observed buoyant mass versus vibrational mode frequency give experimental counterparts to the theoretical curves in Fig.~\ref{figslipDeviation}a. Fitting equation~\eqref{eq:alphaFull} to these measured curves, using a least-squares fitting procedure, produces a single radius and a single slip length (Appendix~\ref{Observed buoyant particle mass using each SMR mode}). Because the expected slip lengths are much smaller than the particle radius, the resulting effect of slip on the frequency shift curves is very small. To overcome this difficulty, the measurement process is repeated hundreds of times on each particle---providing histograms of the measured radius and slip length.

Histograms of the measured radii of individual particles, from repeat measurements, appear to be normally distributed with small variance, e.g., see Fig.~\ref{fighistograms}a. This distribution is consistent with the central limit theorem, and as such, the average of these measured distributions should provide the true particle radius---small bias exists that is negligible relative to the observed standard error; see Supplementary Fig.~1.

In contrast to the measured radii, however, histograms of the measured slip lengths display a distinct right skewness with large variance, e.g., see Fig.~\ref{fighistograms}a. To robustly measure the actual slip length, i.e., without the convoluting effects of measurement noise, we investigate the origin of this difference in histograms for the  particle radius and slip length, on a single particle. This is achieved using Monte Carlo simulations of the measurements that involve numerically synthecized data and encorporate frequency noise levels observed in measurements (Appendix~\ref{Monte Carlo simulations}). Simulations corresponding to measurements in Fig.~\ref{fighistograms}a are reported in Fig.~\ref{fighistograms}b and bear a striking resemblance. This shows that the observed skewness is a direct result of frequency noise and immediately suggests that  skewness in the slip length histograms can be deconvoluted to recover the actual slip length. A theoretical analysis is performed on the fitting procedure to connect the observed slip lengths to the actual slip length (Appendix~\ref{Measuring slip lengths from skewed histograms}). Figure~\ref{fighistograms}c gives the results of this analysis by showing the mean of the slip length histogram versus the standard deviation in the frequency noise (averaged over all measured SMR modes). The derived theory independently predicts the Monte Carlo simulations and enables the actual slip length to be determined.

Particles of radii in the range 115 to 165 nm are used, because particles of smaller size exhibit unacceptable signal-to-noise (frequency shift signal is proportional to radius cubed) while those of larger size are affected by the bounding walls of the SMR used in this study. See Supplementary Fig.~7a for the full dataset including those particles excluded for the reasons stated above. 

The synthesis process for the gold nanoparticles results in deviations from sphericity. The effect of this non-sphericity is fully characterised using both analytical and numerical calculations, alongside TEM images of the nanoparticles; see Supplementary sections 7 and 8. The developed protocol is insensitive to particle non-sphericity that results in a negative slip length bias that is much smaller than, and therefore cannot account for, the measured positive slip length. This bias is one order-of-magnitude smaller than the reported slip lengths and well within the reported 95\% confidence intervals.

\subsection{Measured Navier slip length}

\begin{figure}
\centering
\includegraphics[width=\columnwidth]{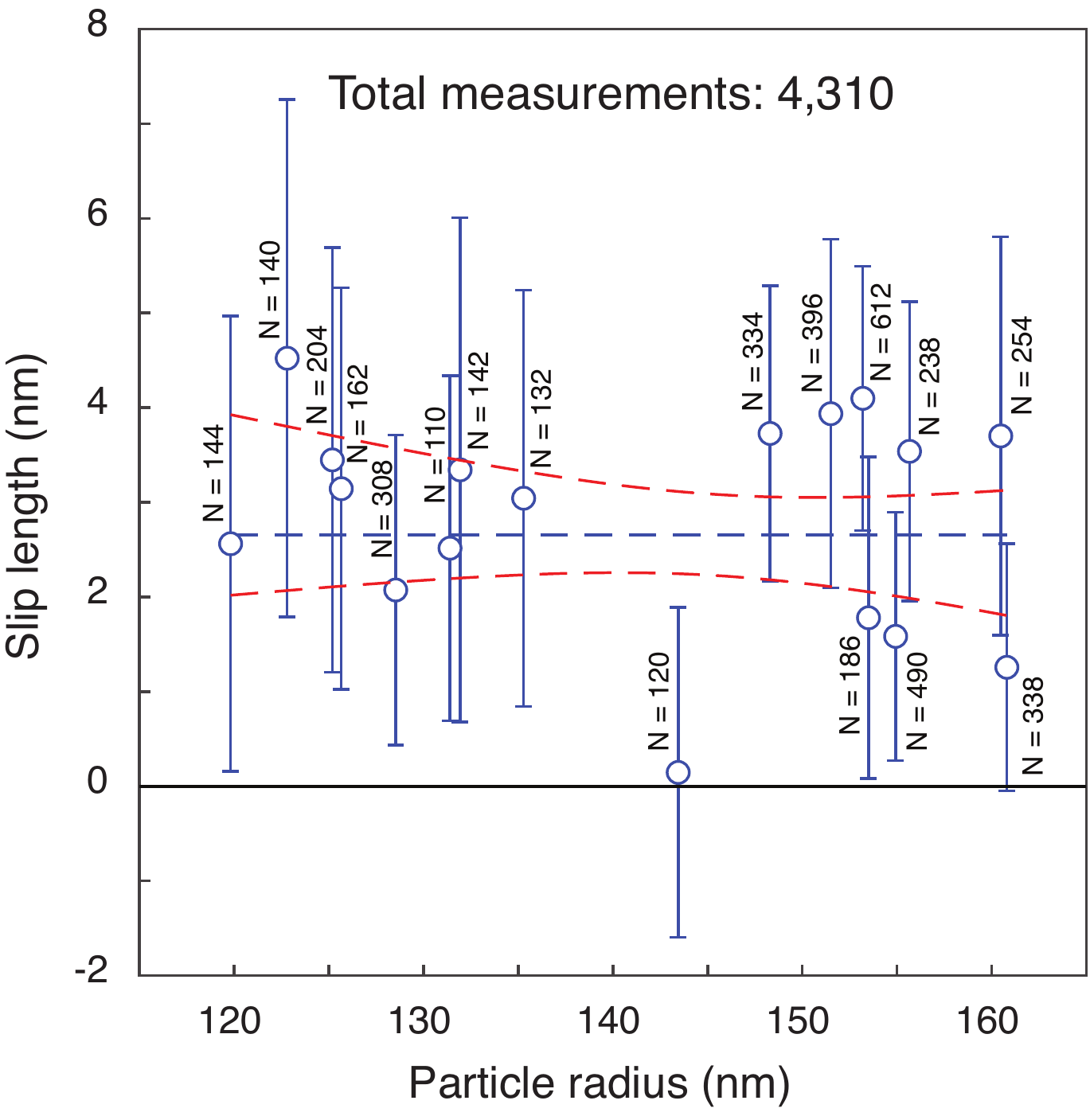}
  \caption{ \rm Measured Navier slip length versus nanoparticle radius for 17 individual gold nanoparticles of different radii---the full particle size range accessible using the present SMR is shown. The complete dataset of 22 particles measured, including smaller and larger particles (radii 67.9--205.0 nm) excluded for unacceptable signal-to-noise or wall effects (discussed above), is given in Supplementary Fig.~7a. Linear regression fit to the measured slip lengths of the remaining 17 particles versus their radii (119.8--160.8 nm)---shown here---is performed; this fit is weighted by the inverse squares of the standard error of each slip length measurement. Red dashed (curved) bands define two standard errors for  straight lines of varying slope that fit the data. The blue dashed horizontal line (zero slope) is the best fit to a constant slip length. The solid black horizontal line corresponds to no-slip---it is provided for reference. Error bars specify a 95\%~C.I., with uncertainties in the particle radii being smaller than the size of the data points.}
\label{figexperiment}
\end{figure}

Figure~\ref{figexperiment} reports data for the actual slip length of each nanoparticle---taken from hundreds of individual measurements---and their associated uncertainties (95\%~C.I.). This dataset evidently displays an average slip length, over all nanoparticles, which is nonzero and positive. A statistical analysis using linear regression of the slip length dataset in Fig.~\ref{figexperiment}, weighted by the inverse of the standard errors squared, is performed to assess the independence of the slip length on the particle radius. This shows that there is no evidence for a correlation between radius and measured slip length (Appendix~\ref{Statistical analysis of slip length versus radius data}).

A regression fit using a constant term only is then applied and yields a slip length of $2.7$$\pm$$0.6$ nm (95\%~C.I. with a p-value less than $10^{-6}$). That is, a nonzero (and positive) slip length of approximately ten molecular diameters of water is accurately measured.
We thus conclude that the reported measurements of slip length, on individual particles of different size, are consistent with the constitutive property of the Navier slip condition: the slip length is independent of the particle, and its size, for the same solid material and liquid.

An upper bound for the liquid shear rate at the particle surface, encountered in measurements, is $10^5 \, \text{s}^{-1}$. Molecular simulations show that nonlinearities in slip occur at shear rates at least 3 orders-of-magnitude larger~\cite{bocquet2007}. 
Molecular simulations~\cite{huang2008} also show that the slip length increases with the degree of  hydrophobicity---no-slip occurs for hydrophilic surfaces and a slip length of no more than 20 nm arises for highly hydrophobic surfaces (contact angles of approximately 140$^{\circ}$). Recent experiments~\cite{valsesia2018} show that citrate-stabilized gold nanoparticle surfaces used in this study are weakly hydrophobic with a contact angle in water of about 95$^{\circ}$. The measured slip length of $2.7$$\pm$$0.6$ nm in this study on individual (weakly hydrophobic) citrate stabilized gold nanoparticles is compatible with molecular simulation data for this level of hydrophobicity; see Fig.~3 of Ref.~\cite{huang2008}.

\section{Conclusions}

The uniqueness of this work lies in the ability to measure the Navier slip length of an individual nanoparticle in an unconfined liquid, with sub-nanometer precision, while directly probing its constitutive nature.
This experimental development, its associated theoretical framework, and the proof-of-principle demonstration on  citrate-stabilized gold nanoparticles in water, enable studies  of other nanoparticle systems that can be used to validate the wealth of molecular simulation data on slip. This can include studies as a function of surface  chemistry, particle crystal structure, liquid properties and temperature. Use of multiple SMRs, of different dimensions, would expand the particle radius range accessible. This would provide valuable experimental insight into physical and chemical mechanisms underlying nanoscale flows.

\section*{ACKNOWLEDGMENTS}

The authors gratefully acknowledge support from an Australian Postgraduate Award, the Australian Research Council Centre of Excellence in Exciton Science (CE170100026), the Australian Research Council Grants Scheme and the Institute for Collaborative Biotechnologies through grant W911NF-09-0001 from the US Army Research Office. We also thank the Koch Institute Swanson Biotechnology Center for technical support, specifically The Peterson (1957) Nanotechnology Materials Core Facility for TEM images of gold nanoparticles.

\appendix

\section{Measurement setup}\label{Measurement setup}

The suspended microchannel resonator (SMR) used in this study is a micro-cantilever with an integrated and embedded U-shaped microfluidtic channel; see Figs.~\ref{figschematic}(b,c). In contrast to conventional resonant mass sensors that are immersed in liquid, the SMR encapsulates the liquid environment inside the cantilever structure. The cantilever itself is contained in an on-chip vacuum chamber and vibrates in its resonant modes. This dramatically reduces viscous losses that would otherwise dominate the noise in mass measurements, thereby leading to extreme  precision in frequency measurement. As a nanoparticle passes through the integrated channel, the cantilever’s mass is transiently altered. This induces a brief, measurable change in the cantilever's resonant frequencies that is proportional to the buoyant mass of the particle. The frequency shift of the cantilever's third flexural mode is depicted in Fig.~\ref{figschematic}c. Because higher-order eigenmodes of the cantilever each have a unique vibrational deflection function (mode shape)~\cite{olcum2015}, the frequency shift signal acquired from each mode is different; see Fig.~\ref{figschematic}d.

The core measurement technique used in this study relies on simultaneously exciting several higher-order eigenmodes of the SMR, while a single nanoparticle is flowed back-and-forth through the SMR. The oscillation amplitudes of all modes are systematically decreased until (nonlinear) mechanical crosstalk between the modes is negligible. Slip at the particle surface affects each vibrational mode differently---due to their different excitation frequencies---and the effect of slip on these independent modes are measured concurrently.

To simultaneously acquire the frequency shift signals of multiple modes, the cantilever is configured as an oscillator with each mode having its own feedback loop; see Fig.~\ref{figschematic}c and Ref.~\cite{olcum2015}. Namely, the tip vibration of the SMR, which involves a superposition of all oscillated mode shapes, is measured using an optical lever setup. A custom circuit with adequate signal bandwidth (10 MHz) is utilized to condition, bandpass filter and amplify the signal acquired from a fast split-photodiode. To maintain linearity in the detected signal and prevent any crosstalk between different resonant modes, the gain of the photodetector is systematically reduced; this also prevents any signal saturation or clipping. The amplified detector signal is fed into a field programmable gate array (FPGA) through an analog-to-digital converter. The FPGA circuit is programmed to maintain an array of digital phase-locked-loops (PLL). Each PLL is dedicated to a single resonant mode and placed in closed-loop with the resonator, both to (i) demodulate the frequency variations of the vibrational mode, and (ii) excite the mode at a constant amplitude, as a particle induces a time-varying change in the cantilever at its resonant frequency. The theory of how a PLL should be configured to excite and de-modulate time-varying resonant frequency changes is discussed elsewhere~\cite{olcum2015}. In contrast to previous studies, this capability is extended up to the $6^\mathrm{th}$ flexural mode of an SMR. During measurements, the loop bandwidth for each PLL controlling a specific mode is set to at least 250 Hz with a sampling rate of 1~kHz. Because each vibrational mode has a different frequency, the PLLs operate independently in the FPGA. Similarly, the actuation signals generated by each PLL are combined by the FPGA and amplified using an RF amplifier driving a piezo-ceramic actuator placed underneath the SMR chip. Multiple modes of the SMR are simultaneously and independently actuated by the ceramic actuator that concurrently vibrates at multiple frequencies; this is because each vibrational mode is responsive to signals only at its own resonant frequency.

To minimize the effects of uncertainty in the particle position and trajectory within the fluidic channel, frequency shifts of the first six resonant modes of the SMR (due to the passage of a single nanoparticle) are  measured simultaneously. Fundamental mode 1 is not used in analysis because it does not contain nodes, which limits the ability to accurately determine particle position (Appendix~\ref{Observed buoyant particle mass using each SMR mode}). As a nanoparticle moves through the SMR, the measured frequency shift traces out the curves described by equation (\ref{eq:freqshiftformula}), i.e., the square of the mode shapes; see Fig.~\ref{figschematic}d. This is used to extract the mass discrepancy parameter, $\alpha$, for each mode, from which the nanoparticle radius and slip length are directly calculated (Appendix~\ref{Observed buoyant particle mass using each SMR mode}).

The requirement of wide and high channels (much larger than the particle size) and ability to control multiple modes simultaneously can increase the mass-equivalent noise in measurements relative to previously reported nanoparticle measurements~\cite{olcum2014}. To circumvent this problem, a microfluidic control method is used that was previously applied to monitor the buoyant mass of a single cell throughout its cell-cycle~\cite{godin2010}. Specifically, the same nanoparticle is passed back and forth through the SMR's fluidic channel by controlling the fluidic pressures on both sides of the SMR. To enable this nanoparticle flow-control, long and separate fluidic channels (10~$\mu$m-wide and 625~$\mu$m-long) are connected to the SMR; see `Trap channels' in Fig.~1b. After each individual particle measurement, the nanoparticle is held in these long channels for 2 seconds before being passed back into the SMR for the next single measurement.

\begin{table}[h]
\footnotesize
\caption{\label{table1}\rm \footnotesize Dimensions of the SMR device used in this study.}
\begin{ruledtabular}
\begin{tabular}{c c}

 Property & Dimension ($\mu$m) \\ 
\hline
 SMR length &  400 \\ 
  SMR width	& 19 \\ 
  SMR thickness	& 4 \\ 
  Channel height	& 3 \\ 
  Channel width	& 5 \\ 
  Lid thickness	& 0.5 \\ 
  Wall width	& 2 \\ 
\end{tabular}
\end{ruledtabular}
\end{table}

\section{Design of the suspended microchannnel resonator}\label{Design of the suspended microchannnel resonator}

Behavior of the mass discrepancy parameter in the limit of small inertia, i.e., $\beta \ll 1$ (viscous dominated flow), is given asymptotically by $\alpha = 1 - (\sqrt{2}/9)\left(\gamma - 1\right)\beta^{3/2} + O(\beta^2)$ whereas in the opposite limit of high frequency, i.e., $\beta \gg 1$ (inviscid flow), we have $\alpha = 3/(1+2\gamma)+ O(1/\sqrt{\beta})$; see equation~(\ref{eq:alphaFull}) and Supplementary section 1 for the full derivation. This establishes that slip has no effect in these limits---slip affects the flow only at intermediate frequency, $\beta$. Therefore, it is important to use multiple vibrational modes of the SMR within this intermediate frequency range---enabling the optimal interrogation of slip. Moreover, to satisfy the underlying theoretical assumption of an unbounded flow, the dimensions (width and height) of the integrated fluidic channel must be much larger than the largest target particle diameter interrogated. To meet these criteria, an SMR with the dimensions and mechanical parameters reported in Table~\ref{table1} is designed, fabricated and used. The SMR is fabricated in the Microsystems Technology Laboratories at MIT, with some steps performed at Innovative Micro Technology, Santa Barbara, CA, utilizing a previously described method~\cite{lee2010}. The resonance properties of flexural modes 2 to 6 of the SMR are provided in Table~\ref{table2}. Fundamental flexural mode 1 (which is not used) has a resonant frequency of 34.796 kHz and quality factor of 20,761.

\begin{table}[h]
\footnotesize
\caption{\label{table2}\rm \footnotesize Properties of flexural vibrational modes of the SMR detailed in Table~\ref{table1}.}
\begin{ruledtabular}
\begin{tabular}{c c c c}
Mode & Unloaded & Quality & Standard deviation of \\
number & frequency (MHz) & factor & frequency noise (Hz) \\
\hline
2 & 0.2180 & 9,440 & 0.43 \\
3 & 0.6094 & 4,773 & 0.37 \\
4 & 1.191 & 2,823 & 0.35 \\
5 & 1.963 & 2,093 & 0.61 \\
6 & 2.921 & 1,620 & 1.6
\end{tabular}
\end{ruledtabular}
\end{table}

\section{Observed buoyant particle mass using each SMR mode}\label{Observed buoyant particle mass using each SMR mode}

Experimental data for the frequency shift time series (e.g., see Fig.~\ref{figschematic}d) is fit to equation~(\ref{eq:freqshiftformula}) using an arbitrary polynomial (up to $3^\mathrm{rd}$ order) for the particle position, $z_p$, versus time, $t$, measurement. This fit procedure enables a nonlinear dependence of $z_p$ on $t$ to be accurately described---it is approximately linear. A least-squares method is applied to each mode and is used to simultaneously determine (i) the required mass ratio, $M^\mathrm{obs}_\mathrm{p}/M_\mathrm{SMR}$, in equation~(\ref{eq:freqshiftformula}), and (ii) the constants in the above-described ($z_p$ vs $t$) polynomial function.

\textit{Calibration of SMR mass.} To measure the observed buoyant mass of the particle, $M^\mathrm{obs}_\mathrm{p}$, the mass of the SMR, $M_\mathrm{SMR}$, is required. This is determined using NIST-tracable polystyrene particles (ThermoFisher 4016A) of known mass. These particles have a well characterized radius of 793.5$\pm$9(SD) nm and a density of 1,050~kg/m$^3$. A total of 341 particles are measured using the SMR from a random sample of these NIST particles. The resulting measured radius of each particle is taken as the average of two measurements: the particle travels through each arm of the SMR's microfluidic channel as it traverses from one `buried channel' to the other (on the other side of the SMR); see Fig.~\ref{figschematic}c. The temperature is monitored to be between 22.4 to 23.0$^\circ$C throughout the calibration procedure. Because these polystyrene particles are much lighter than the gold particles used in the slip measurements, their motion relative to the SMR's solid walls is small. As such the effects of slip are minimal; this is evident from equation~(\ref{eq:alphaFull}) in the limit, $\gamma \rightarrow 1$.

Equations~(\ref{eq:freqshiftformula})--(\ref{eq:alphaFull}) with $b = 0$ (no-slip) are used to analyze the 341 observed mass measurements of the NIST particles, producing a  histogram of the normalized particle mass, $M^\mathrm{obs}_\mathrm{p}/M_\mathrm{SMR}$. The SMR's mass, $M_\mathrm{SMR}$, is then chosen such that the mean of this histogram matches the mean of the NIST specified particle mass. This results in an SMR mass of $M_\mathrm{SMR} = 5.387$$\pm$$0.026 \times 10^{-11}$ kg, where the reported 95\% C.I. is the combination of (i) the uncertainty specified by two standard errors of the measured mean particle radii, and (ii) the uncertainty due to temperature variation. This value for $M_\mathrm{SMR}$ is used in all gold nanoparticle measurements that interrogate slip.

To explore sensitivity to the (no-slip) assumption of $b = 0$ used in these calibration measurements, the same procedure is applied with $b = 10$ nm and $b = 100$ nm where the SMR mass is determined to be $5.386\times 10^{-11}$ kg and $5.377\times 10^{-11}$ kg, respectively---a change in SMR mass of only $0.02\%$ and $0.2\%$, respectively, which is smaller than the reported uncertainty in $M_\mathrm{SMR}$ above. This shows that the hydrodynamic flow generated by these particles, and associated boundary condition at the polystyrene-water interface, exert a negligible effect on the measured SMR mass and can be safely ignored.

\textit{Gold nanoparticle measurements.} Details of the gold nanoparticles used in measurements are in Appendix~\ref{Selecting appropriate gold nanoparticles}. For each of the five measured SMR modes (modes 2 to 6, see Fig.~\ref{figslipDeviation}b), the observed buoyant mass of each gold nanoparticle is  determined using equation~(\ref{eq:freqshiftformula}). Typical frequency changes due to the presence of a nanoparticle are approximately 1 to 10 Hz and always greater than the standard deviation of the frequency noise (which enables detection of the required signal). Equation~(\ref{eq:alphaformula}) then gives the value of $\alpha M_\mathrm{p}^\mathrm{true}$, where the true buoyant mass of the particle is $M_\mathrm{p}^\mathrm{true} = (4\pi/3)R^3(\rho_\mathrm{p} - \rho_\mathrm{f})$; $\rho_\mathrm{p}$ is the known particle density, i.e., density of gold, and $\rho_\mathrm{f}$ is the known liquid density, i.e., density of water at the measured temperature (between 22.4 and 23.4$^\circ$C for the gold nanoparticle measurements). The mass discrepancy parameter, $\alpha$, is a function of the unknown particle radius, $R$, and slip length, $b$; see equation~(\ref{eq:alphaFull}). A nonlinear least-squares approach is used with equation~(\ref{eq:alphaFull}) to extract the two required parameters, $R$ and $b$, from the (five independently measured) mass versus frequency data of the same particle (one measured mass for each vibrational mode, i.e., SMR modes 2 to 6). For the present experiments, slip lengths of a few nanometers will alter the measured frequency shift curves by approximately 0.01--0.1~Hz, relative to the no-slip result. Because these changes in the frequency shifts due to slip are expected to be much smaller than the standard deviation of the frequency noise (on the order of 1Hz, see Table 2), each nanoparticle measurement is repeated hundreds of times; see Table~\ref{table3}. Cumulative histograms of the radius and slip length for an individual nanoparticle are then generated; see Fig.~3a for an example of these distributions.

\begin{table}[h]
\footnotesize
\caption{\label{table3}\rm \footnotesize Measured radius, slip length and number of measurements on each gold nanoparticle. Reported uncertainty is two standard errors of the mean.}
\begin{ruledtabular}
\begin{tabular}{c c c}
Radius (nm) & Slip length (nm) & Number of \\ & & measurements \\
\hline
119.8$\pm$0.4 & 2.6$\pm$2.4 & 144 \\
122.8$\pm$0.4 & 4.5$\pm$2.7 & 140 \\
125.2$\pm$0.4 & 3.4$\pm$2.2 & 204 \\
125.6$\pm$0.4 & 3.1$\pm$2.1 & 162 \\
128.5$\pm$0.3 & 2.1$\pm$1.6 & 308 \\
131.4$\pm$0.4 & 2.5$\pm$1.8 & 110 \\
131.9$\pm$0.6 & 3.3$\pm$2.7 & 142 \\
135.3$\pm$0.5 & 3.0$\pm$2.2 & 132 \\
143.4$\pm$0.5 & 0.2$\pm$1.7 & 120 \\
148.3$\pm$0.4 & 3.7$\pm$1.6 & 334 \\
151.5$\pm$0.4 & 3.9$\pm$1.8 & 396 \\
153.2$\pm$0.3 & 4.1$\pm$1.4 & 612 \\
153.5$\pm$0.4 & 1.8$\pm$1.7 & 186 \\
154.9$\pm$0.3 & 1.6$\pm$1.3 & 490 \\
155.7$\pm$0.4 & 3.5$\pm$1.6 & 238 \\
160.5$\pm$0.4 & 3.7$\pm$2.1 & 254 \\
160.8$\pm$0.4 & 1.3$\pm$1.3 & 338
\end{tabular}
\end{ruledtabular}
\end{table}

\section{Selecting appropriate gold nanoparticles}\label{Selecting appropriate gold nanoparticles}

 Stabilized suspensions of gold nanoparticles in citrate buffer from Sigma-Aldrich (742066, 742074, 742082 and 742090) are used for all slip measurements. These nanoparticles have identical composition and surface chemistry, but vary in size. The developed theory in \S\ref{secTheory} assumes a rigid, spherical particle of constant density. However, the true composition of these particles features a citrate layer, of thickness ($<\!1$ nm), adsorbed to the particle surface~\cite{park2014}.
Because these layers are of low density, soft and compliant, they must produce an even smaller negative bias in the measured slip length, i.e., they cannot account for the measured positive slip length reported in Fig.~\ref{figexperiment}. These citrate layers do modify the wettability of the  gold surface, which affects the slip length; this property is used to compare the measured slip length to molecular simulations. The nanoparticles are sampled from populations with mean radii ranging from 100 to 200 nm.

To ensure only a single gold nanoparticle is in the channel at any given time, we use the following procedure. First, a low particle concentration of 10$^4$ to 10$^5$ particles per ml is supplied to the SMR channel. Second, we use the frequency versus time curves acquired from the multimode measurements (e.g., see Fig.~\ref{figschematic}d) to test for the presence of more than one particle. If two particles are in the SMR simultaneously, a convolution of two different signals is measured; this looks drastically different to a single particle signal; e.g., see Fig.~4 of Ref.~\cite{olcum2015}. Occasional measurements of this type are discarded.

A total of 22 gold nanoparticles are measured (hundreds of times each, see above) whose radii vary from 67.9 to 205.0 nm. Slip lengths of a subset of these particles are interrogated, for the following reasons. Reducing the particle size lowers the signal-to-noise ratio, restricting the practical minimum radius that can be measured to 115 nm; see Supplementary Figs.~3 and 4a. Conversely, the derived theory implicitly assumes the particle does not interact hydrodynamically with the bounding solid walls of the SMR, leading to a practical upper limit on the particle radius of 165 nm (above which, leads to errors in the measured slip length of greater than 1 nm); see Supplementary section 6. The full dataset (including those particles not analysed as discussed above) is given in Supplementary Fig.~7a.

17 individual nanoparticles lie within this particle size range allowed by the measurement setup: radii between 115 and 165 nm. Use of these particles thus permits a robust assessment of the constitutive nature of the Navier slip condition. An analysis of the various uncertainties is  provided next; error bars derived from this analysis are included in the data reported.

\textit{Sources of uncertainty.} There are three key sources of uncertainty in the particle measurements: (i) a finite number of particle measurements leads to inevitable uncertainty in the measured means of the histograms, (ii) the temperature is measured to vary between 22.4 and 23.4$^\circ$C (for the gold nanoparticle measurements) which alters the density and viscosity of the liquid, and (iii) measurement of the cantilever mass has uncertainty detailed in Appendix~\ref{Observed buoyant particle mass using each SMR mode}. All these uncertainties can be quantified and are independent, so the total uncertainty is the RMS of these values; the reported uncertainties in Table~\ref{table3} give two standard errors of the mean.

\section{Effect of particle non-sphericity on slip length measurements}\label{Effect of surface roughness on slip length measurements}
Due to their synthesis process, gold nanoparticles are not perfectly spherical---but the measurement protocol is insensitive to  non-sphericity, as we now show. Non-sphericity is studied by superposing a shape perturbation function, $g$, on the surface of a perfect sphere; the $O(1)$ function, $g$, is dimensionless. The radial coordinate of the particle surface is,
\begin{equation}
r = 1 + \xi g(\theta, \phi),
\end{equation}
where $r$ has been non-dimensionalised by the radius of an equivalent volume sphere, $\xi$ is the RMS surface roughness and $\theta$ and $\phi$ are the usual spherical polar and azimuthal angles, respectively. The nanoparticles assume a random orientation in the SMR and hundreds of measurements are taken on each nanoparticle. Therefore, it is appropriate to study the ensemble average effect of non-sphericity over all possible particle orientations. In Supplementary section 7, we prove that this averaging procedure---inherent in our measurements---extracts the radius of an equivalent volume sphere to $O(\xi)$.

Numerical simulations of non-spherical particles, using finite element analysis, shows that there exists a bias to the slip length at $O(\xi^2)$; see Supplementary section 8 for the complete analysis. Critically, this produces a \textit{negative} slip length bias and is one order-of-magnitude smaller than the positive slip lengths reported here (bias is $-0.18$$\pm$$0.03$ nm; 95\% C.I.); it cannot account for the measurements. Note that this bias is often termed an `effective slip length' in the literature~\cite{kamrin2010}.

\section{Effect of the SMR walls on slip length measurements}\label{Effect of the SMR walls on slip length measurements}
The disturbance velocity field created by the particle (when inside the SMR) is inviscid outside the particle's viscous boundary layer. In this outer region, it decays as $1/r^3$ where $r$ is the radial distance from the particle center. For the present experiments, the viscous boundary layer thickness is between 200 and 800 nm, which is smaller than the channel width and height. This rapid decay in the disturbance flow minimizes the effect of the SMR walls on the measurements.

To quantify wall effects on the slip measurements, numerical simulations are performed for finite channel size. A boundary integral method (described in Supplementary section 5) is used. Simulations are performed on no-slip particles to create sample data from which a radius and slip length is extracted. The same procedure used for the experiments as described in Appendix~\ref{Observed buoyant particle mass using each SMR mode} is employed here. A fitted slip length of zero indicates the bounding SMR walls have no effect on the measurements, while a nonzero fitted slip length indicates a measurement bias. While these simulations of wall bias use the no-slip boundary condition, they apply equally to particles with slip by linearity (provided the slip length is much smaller than the particle radius, which is the experimental situation).

Transit times (which vary between 300 and 800 ms for each half of the SMR) are used to estimate the particle's proximity to the SMR walls. In general, particles close to the SMR walls have a large negative bias to the slip length while particles far from any walls do not experience any bias. From the measured transit times, we conclude that the vast majority of particles do not lie close to the internal walls. The primary finding of our numerical simulations is that particles of radius smaller than 160 nm induce a negligibly small bias in slip (due to the walls) while particles of radius close to 200 nm experience a detectable negative bias in the extracted slip length. Therefore, these larger particles are excluded from the final analysis; see Supplementary Fig.~7 for the full dataset, including these excluded particles. Supplementary section 6 gives further details on the analysis of the wall effects.

Analyses presented here and in the preceding section show that experimental non-idealities---involving particle non-sphericity and SMR wall effects---produce small and negative slip length biases only, Their presence cannot explain the measured positive slip lengths in Fig.~\ref{figexperiment}.

\section{Monte Carlo simulations}\label{Monte Carlo simulations}

Monte Carlo simulations are performed where synthetic frequency shift data are generated using equation~(\ref{eq:freqshiftformula}), for a chosen (nominal) particle mass, radius, density ratio and slip length. Gaussian frequency noise with the same standard deviation  found in the measurements of each SMR mode (see the fourth column in Table \ref{table2}) is added to this synthetic frequency shift data{; this produces simulated data resembling the curves in Fig.~\ref{figschematic}d. The particle radius and slip length are then recovered from this noisy synthetic data using an identical procedure to that of the measurements, i.e., a least squares procedure is used with equations (\ref{eq:freqshiftformula})--(\ref{eq:alphaFull}) to determine $R$ and $b$ from modes 2 to 6 (Appendix~\ref{Observed buoyant particle mass using each SMR mode} for a detailed description).

In total, 10,000 simulations are performed using this procedure on a single particle specification, resulting in histograms for both the extracted particle radius and slip length. Here, a particle radius and slip length of 125 nm and 5 nm, respectively, is used in the simulations to test the efficacy of the data fitting procedure; simulations varying both the particle radius and slip length are given in Supplementary section 3. The resulting radius and slip length histograms for these simulations are provided in Fig.~\ref{fighistograms}b.

The histogram for the radius appears to be normally distributed with a mean of 125.03$\pm$0.06 nm, i.e., the particle radius of 125 nm is extracted accurately. Additionally, the variance of this histogram matches the experimental data, demonstrating that frequency noise is the direct cause of the variance in the experimental particle radius distributions. In contrast, the slip length histogram is right skewed, as in the experimental data. The variance and the skewness coefficient match the experimental data, again indicating the shape of these distributions is a direct consequence of frequency noise. The mean of the slip length distribution is 6.91$\pm$0.24 nm which exceeds the specified slip length of 5 nm; this discrepancy is now discussed.

To explore the difference between the specified and extracted mean slip lengths, as a function of frequency noise, we run a set of Monte Carlo simulations where the frequency noise is systematically increased from zero (all other details are as described above). This is achieved by fixing the relative strengths of the frequency noise standard deviations across all vibrational modes  to the experimental situation, and increasing their magnitudes; Fig.~3c reports the results of these simulations. Because the mass discrepancy parameter, $\alpha$, is proportional to the frequency shift of each mode, $\Delta f_n$ (see equations~(\ref{eq:freqshiftformula}) and (\ref{eq:alphaformula})), adding noise to the frequency shift curves is equivalent to adding noise directly to the mass discrepancy parameter, $\alpha$. We therefore report the average standard deviation of the mass discrepancy parameters over all modes on the horizontal axis in Fig.~3c.

The mean slip length (extracted from the slip length distributions) appears to increase quadratically with increasing noise, suggesting the actual slip length can be determined if the noise level is known. This is discussed next in Appendix~\ref{Measuring slip lengths from skewed histograms}.

\section{Measuring slip lengths from skewed histograms}\label{Measuring slip lengths from skewed histograms}

To extract the actual slip length from experimental and synthetic data, an asymptotic analysis---in the limit of small frequency noise---is performed on the least-squares fit procedure used to determine $R$ and $b$. That is, the primary (leading-order) effect of skewness in the histograms is considered. This gives a formula connecting (i) the \textit{measured mean slip length and particle radius} of each dataset and (ii) the frequency noise of each SMR vibration mode, to the \textit{actual slip length and particle radius} corresponding to the Navier slip condition. The full derivation is given in Supplementary section 2. This procedure is validated using Monte Carlo simulations with frequency noise of identical magnitude to measurements (Appendix~\ref{Monte Carlo simulations} and Fig.~3c). By performing a large number of measurements on an individual particle, the noise can be well characterised allowing for the recovery of the actual slip length.

\section{Statistical analysis of slip length versus radius data}\label{Statistical analysis of slip length versus radius data}
A statistical analysis using linear regression is performed to test for independence of the slip length on particle radius. A high p-value ($\mathrm{p}=0.57$) is observed indicating there is no evidence for a correlation between radius and measured slip length. Moreover, the R$^2$ value of this linear regression is very small 0.018 ($\ll$1). Hypothetically, if the p-value were to be small (e.g., less than 0.05), this minute R$^2$ value would indicate that the radius accounts for 2\% of the observed variance in the measured slip length---which is also negligible. Thus, regardless of the p-value there is no statistically significant and meaningful relationship between the measured slip length and particle radius. A similar conclusion arises if a higher-order polynomial is used. 

\newpage

\def\bibfont{\footnotesize}

\newpage

\end{document}